\documentclass[%
 reprint,
 superscriptaddress,
 amsmath,amssymb,
 aps,
]{revtex4-2}

\usepackage{graphicx}% Include figure files
\usepackage{dcolumn}% Align table columns on decimal point
\usepackage{bm}% bold math
\usepackage{hyperref}% add hypertext capabilities
\DeclareMathOperator{\sech}{sech}

\usepackage{xcolor}

\begin{document}
\preprint{APS/123-QED}

\title{Reverse design of the ideal pulse for hollow capillary fiber post-compression schemes}

\author{Marina Fern\'andez Gal\'an} \email{marinafergal@usal.es}
\author{Enrique Conejero Jarque}
\author{Julio San Roman}
\affiliation{Grupo de Investigaci\'on en Aplicaciones del L\'aser y Fot\'onica, Departamento de F\'isica Aplicada, Universidad de Salamanca, E-37008 Salamanca, Spain}
\affiliation{Unidad de Excelencia en Luz y Materia Estructuradas (LUMES), Universidad de Salamanca, Salamanca, Spain}

\date{\today}

\begin{abstract}
The countless applications of ultrashort laser pulses in very different scientific areas explain the ongoing efforts to develop new strategies for the generation of light pulses with increasingly better characteristics. In this work, we theoretically study the application of the nonlinear reverse propagation method to produce few-cycle pulses with clean temporal profiles from standard post-compression setups based in hollow capillary fibers. By numerically solving the propagation of a desired goal pulse in the backward direction, we are able to predict the structure of the ideal input pulse that could be perfectly compressed in a given setup. Although the goal pulse cannot be chosen in a
simple manner due to the fundamental symmetries of the nonlinear propagation equation, our analysis shows that the ideal pulse presents a recurring form and that, in general, both its intensity profile and phase must be shaped to recover the optimized goal output.
\end{abstract}

%\keywords{Suggested keywords}
\maketitle
%\tableofcontents

%--------------------------------------------------------
\section{Introduction}

In the last decades, the rapid growth of ultrafast science demanding ultrashort laser pulses has pushed the community to develop different compression schemes for the generation of tailored light pulses with optimal characteristics \cite{Nagy2021}. These pulses have broken new ground in time-resolved spectroscopy \cite{Maiuri:2020}, and have become a key enabling tool for strong field physics leading to the generation of the shortest attosecond pulses \cite{Corkum2007}, which offer great promise for advancing precision control of ultrafast dynamics at the atomic scale \cite{Goulielmakis2010,Nisoli2017,Shi2020,Borrego-Varillas2022}. Among the main properties of ultrashort laser pulses, their temporal duration is the one that is most often desired to be pushed to the limit. The standard technique which is routinely used by many laboratories to temporally compress intense light pulses in the optical spectral region towards their ultimate duration relies on the nonlinear propagation through gas-filled hollow capillary fibers (HCFs), as first proposed by Nisoli and coworkers in 1996 \cite{Nisoli1996}. This method is based on the nonlinear spectral broadening of an input femtosecond pulse by self-phase modulation (SPM) in a HCF, followed by phase compensation in an external compressor comprising dispersive elements such as chirped mirrors, prisms or gratings \cite{Jarque2018}. Intense near-single-cycle pulses have been obtained using this post-compression scheme in combination with positive pressure gradients and a d-scan compressor \cite{Bohle2014}, or with high third-order dispersion (TOD) to group-delay dispersion (GDD) ratio materials such as water \cite{Silva2014,Fabris2015} or ammonium dihydrogen phosphate (ADP) \cite{Timmers2017}, or by tuning the spectral phase of the input pulse and using broadband chirped mirrors \cite{Silva2018}.

Another promising alternative for the generation of ever shorter sub-cycle optical pulses comes through high-energy soliton self-compression in HCFs \cite{Travers2019}, which has been demonstrated for short pre-compressed pulses ($\sim$10 fs at 800 nm) \cite{Travers2019,Brahms2020}, and also proposed for standard multi-cycle pulses in fibers with a decreasing pressure gradient \cite{Galan2022,Galan2023}. Although all these pulses have extremely short temporal durations, most of them also present secondary structures accompanying the main ultrashort peak. Thus, there is still room for improvement and it is natural to ask whether these structures could be cleaned, preferably by shaping the less-sensitive input pulse. However, the complex nonlinear propagation dynamics often results in dramatic pulse reshaping, challenging any explicit design of the initial pulse. In this work, we have addressed this problem and found a positive answer, at least from a theoretical point of view, in an attempt to provide theoretical support and guidelines for in-line HCF-based post-compression experiments that use input pulse shaping. Our strategy is based on the nonlinear reverse propagation method, which has been already applied in the context of highly nonlinear regimes.

The reverse design method undoing all dispersion and nonlinear effects for output pulse optimization in an optical fiber was first proposed in 2003 \cite{Tsang2003}. This work demonstrated that it is possible to theoretically predict the exact input pulse shape that yields a desired output by numerically solving the nonlinear propagation equation backwards, showing that the resulting optimization is much better than that achieved with other techniques such as optical phase conjugation \cite{TsangCONJ2003}. Nonlinear reverse propagation in optical fibers was later used to generate optimally shaped ultrashort pulses relevant for coherent control and for selective two-photon excitation of dye molecules \cite{PawlowskaOL2012,Pawlowska2012}, proving to be a superior method than linear dispersion compensation. More recently, backward propagation has been demonstrated in highly nonlinear regimes like filamentation \cite{Berti2014}, which has unexpectedly proven to be reversible in many cases of interest despite intensity clamping, ionization induced nonlinear losses and extreme phase sensitivity after beam collapse \cite{Shim2012}. In this context, the nonlinear reverse propagation method has also been used to design the initial waveforms which yield a required complex target electric field at a remote distance after undergoing filamentation \cite{Berti2015,Zou2022}. These previous works constitute a clear evidence that the actual reversibility of real physical systems can expand beyond the domain of formally well-posed inverse problems, opening the way to explicit pulse design for control and optimization.

In this paper we study the application of nonlinear reverse propagation to find the ideal input pulse that could lead to a perfect compression in a HCF-based post-compression scheme, directly yielding a clean few-cycle pulse. We first briefly discuss the fundamental symmetries of the nonlinear pulse propagation equation and its implications for reverse design. We also note that, despite HCFs present important guiding losses, the reverse design method can still be used, although it has to be applied carefully. As explained in \cite{Zou2022}, we also find that, in this nonlinear regime, the reverse propagation method requires an intermediate state, which in our case corresponds to the output pulse from the fiber before undergoing perfect phase compensation in a standard compressor (gratings or d-scan). On the whole, our method predicts the intensity profile and phase of the ideal pulse that could be perfectly post-compressed after propagation through the HCF and the compression stage.  Most noticeably, the ideal pulse spectrum always exhibits a similar modulated structure, typically consisting of a main central peak and a pair of side lobes, independent of the chosen setup. Finally, we demonstrate that, in general, both the ideal amplitude and phase distributions are necessary to achieve the desired compressed pulse, revealing the intricate sensitivity of the nonlinear propagation equation to the initial condition.

%--------------------------------------------------------
\section{Methods}

\subsection{Nonlinear propagation of ultrashort laser\\ pulses in hollow capillary fibers}

In the low-intensity regime, where the peak power and the peak intensity of an ultrashort laser pulse remain, respectively, below the critical power for self-focusing and the threshold intensity for gas ionization, the nonlinear propagation through a gas-filled HCF can be accurately modeled with the one-dimensional generalized nonlinear Schrödinger equation (GNLSE) \cite{Nurhuda2003}. In this approximation, the transverse pulse profile is assumed to remain unmodified along the propagation and equal to a waveguide mode. The propagation equation for the complex temporal envelope $A(z,T)$ of a linearly polarized electric-field can then be written in the frequency domain as \cite{Agrawal}:
\begin{eqnarray}
    \frac{\partial\Tilde{A}}{\partial z}+\frac{\alpha(\omega)}{2}\Tilde{A}-i\left[\beta(\omega)-\beta_0-\beta_1(\omega-\omega_0)\right]\Tilde{A} = \nonumber\\ = \mathcal{F}\left\{\hat{N}[A(z,T)]A(z,T)\right\}, \label{eq:GNLSE}
\end{eqnarray}
where $\mathcal{F}$ stands for direct Fourier transform, $\Tilde{A}(z,\omega)=\mathcal{F}\{A(z,T)\}$, $z$ is the propagation coordinate along the fiber, and $T$ is the local time measured in a reference frame traveling with the pulse at the group velocity $v_g=1/\beta_1$. The left-hand side of Eq.~\eqref{eq:GNLSE} includes the effects of linear losses through the absorption coefficient $\alpha(\omega)$ of the HCF, and the complete chromatic dispersion through the frequency-dependent propagation constant $\beta(\omega)$. The parameters $\beta_0$ and $\beta_1$ are defined, respectively, as $\beta_0=\beta(\omega_0)$ and $\beta_1=(d\beta/d\omega)_{\omega_0}$ with $\omega_0$ being the central frequency of the input pulse. On the right-hand side, the operator $\hat{N}[A(z,T)]$ gathers all the nonlinear effects which influence the pulse propagation, and can be expressed in the time domain as:
\begin{eqnarray}
    \hat{N}[A(z,T)]A(z,T) = i\gamma\left(1+\frac{i}{\omega_0}\frac{\partial}{\partial T}\right) \nonumber\\ \times\left(A(z,T)\int_{-\infty}^{+\infty}R(\tau)\big|A(z,T-\tau)\big|^2 \, d\tau\right),
\end{eqnarray}
with the nonlinear parameter given by $\gamma=n_2\omega_0/(cA_{\mathrm{eff}})$, $n_2$ being the nonlinear refractive index of the gas filling the HCF, $c$ the speed of light in vacuum, and $A_{\mathrm{eff}}$ the effective modal area as defined elsewhere \cite{Agrawal}. The operator $(i/\omega_0)\partial/\partial T$ accounts for pulse self-steepening, and the nonlinear response function
\begin{eqnarray}
    R(T) = (1-f_R)\delta(T) \nonumber\\ +f_R\Theta(T)\frac{\tau_1^2+\tau_2^2}{\tau_1\tau_2^2}\exp\left(-\frac{T}{\tau_2}\right)\sin\left(\frac{T}{\tau_1}\right)
\end{eqnarray}
includes both an instantaneous electronic and a delayed molecular contribution to the Kerr nonlinearity in relative fractions $(1-f_R)$ and $f_R$, respectively. Here $\delta(T)$ represents the Dirac delta function and $\Theta(T)$ is the Heaviside step function. The first term in $R(T)$ describes the effect of SPM and the second is responsible for intrapulse stimulated Raman scattering (SRS), which is modeled in a damped harmonic oscillator approximation with two characteristic time constants $\tau_1$ and $\tau_2$ \cite{Couairon2007,Berge2007}.

To solve Eq.~\eqref{eq:GNLSE} numerically, we have implemented the efficient fourth-order Runge-Kutta in the interaction picture algorithm \cite{Hult2007}, in combination with the local error method for continuously adapting the step size \cite{Heidt2009}. For the propagation constant and the absorption coefficient that appear in Eq.~\eqref{eq:GNLSE} we have used the analytical expressions for the fundamental mode of a hollow dielectric capillary, EH$_{11}$, which are given by \cite{Marcatili1964}:
\begin{eqnarray} 
    \beta(\omega)=\frac{\omega}{c}n_{\mathrm{eff}}(\omega)=\frac{\omega}{c}\sqrt{n^2_{\mathrm{gas}}(\omega)-\frac{u_{11}^2 c^2}{\omega^{2}a^{2}}}, \label{eq:beta}\\[2ex]
    \alpha(\omega)=\left(\frac{u_{11}}{n_{\mathrm{gas}}(\omega)}\right)^{2}
    \frac{c^2}{\omega^2a^{3}}\frac{\nu^{2}(\omega)+1}
    {\sqrt{\nu^{2}(\omega)-1}}, \label{eq:alpha}\\
    \text{with} \quad \nu(\omega)=\frac{n_{\mathrm{clad}}(\omega)}{n_{\mathrm{gas}}(\omega)}, \nonumber
\end{eqnarray}
where $u_{11}=2.4048$, $a$ represents the fiber core radius, $n_{\mathrm{clad}}$ is the refractive index of the dielectric cladding and $n_{\mathrm{gas}}$ is the refractive index of the filling gas \cite{Borzsonyi2008}. In our simulations, we have modeled a 3 m long, 125 \textmu m core radius fused-silica HCF filled with argon, with no SRS contribution ($f_R=0$) and a nonlinear refractive index of $n_2 = 1.08 \times 10^{-19}p$ (cm$^2$/W)\cite{Wang2013}, $p$ being the gas pressure in bar.

Given an input pulse $A(0,T)$, Eq.~\eqref{eq:GNLSE} describes its forward nonlinear propagation towards the fiber output in $z=L$. In principle, one could also think of using the GNLSE to recover the original initial condition from the output $A(L,T)$ by reversing the pulse propagation. This can be accomplished by changing $dz$ to $-dz$ in the GNLSE \cite{Tsang2003}, as $z$ is the evolution variable of Eq.~\eqref{eq:GNLSE}. However, when energy losses are taken into account, the resulting backward equation is mathematically ill-posed, meaning that the reverse propagation solution is highly sensitive to small perturbations of its initial condition and to the amplification of noise \cite{Berti2014,Sagiv2020}.

In particular, Eq.~\eqref{eq:GNLSE} includes two different sources of energy dissipation, namely, the linear absorption of the HCF and the non-resonant losses associated with SRS, which transfers part of the pulse energy to the gas molecules due to the interaction with excited rotational states \cite{Agrawal,Blow1989,Mamyshev1990,Bonetti2020}. The latter are always finite and do not represent a problem when simulating the reverse propagation of a laser pulse, because light is never completely absorbed by the molecular medium. However, special care has to be taken when including the fiber losses which, being turned into an exponential gain in backward propagation, can amplify unphysical numerical noise leading to the blow up of the solution. Fortunately, the problematic large-loss regions of a HCF are not within the transparency windows of both the dielectric cladding and the filling gas, so the GNLSE can be numerically reversed with minimal error if the curve $\alpha(\omega)$ as given in Eq.~\eqref{eq:alpha} is restricted to the spectral region of interest, as shown in the Supplemental Material \footnote{See Supplemental Material at [URL] for a more detailed analysis of the numerical reversibility of the nonlinear propagation equation, its fundamental symmetries and the application of the reverse design method to ever shorter near-single-cycle goal pulses. This file also includes additional Refs. \cite{Malitson1965,Tan1998,Loriot2009,Carpeggiani2020,Fisher1983,Sulem}.}. For the simulations performed in the 125 \textmu m core radius HCF filled with Ar at 1 bar, the absorption coefficient was limited to the range of wavelengths shorter than 4.5 \textmu m, where it safely remains below 22 dB/m. Across the full spectral bandwidth of the broadest pulses analyzed, $\alpha(\omega)$ varies smoothly between 2.1 dB/m at 1.5 \textmu m and 0.24 dB/m at 500 nm, not representing a problem for numerical reverse propagation.

\subsection{\label{sec:simetrias}Fundamental symmetries of the\\ propagation equation}

If the GNLSE is written in the time domain by expanding the propagation constant $\beta(\omega)$ in a Taylor series around the central frequency and then applying inverse Fourier transforms (see the Supplemental Material or Refs. \cite{Agrawal,Boyd}), it is straightforward to prove that, in the absence of linear losses ($\alpha=0$) and SRS ($f_R=0$), Eq.~\eqref{eq:GNLSE} is parity-time ($\mathcal{PT}$) symmetric. In other words, it is invariant under the joint transformations of space reversal ($z \rightarrow -z$), time reversal ($T \rightarrow -T$) and complex conjugation ($i \rightarrow -i,\, A \rightarrow A^*$). As a consequence, in this loss-free regime, if $A(z,T)$ is a solution of the propagation equation with initial condition $A(0,T)$, then so is $A^*(-z,-T)$ with initial condition $A^*(0,-T)$. Therefore, as the backward propagation equation is defined as the spatial inversion of the GNLSE, these symmetries imply that, in the absence of linear losses and SRS, the one-dimensional nonlinear reverse propagation of a pulse is equivalent to the forward propagation of its complex conjugated temporal inversion, meaning that both propagation directions are physically indistinguishable \cite{Sagiv2020}.

Furthermore, from the properties of the Fourier transform, it is simple to prove that combined complex conjugation and time reversal of an envelope are equivalent to phase conjugation in the frequency domain \cite{Tsang2003,Lam2021}. Therefore, $\Tilde{A}(z,\omega)$ and $\Tilde{A}^{*}(-z,\omega)$ are the two equivalent solutions of the loss-free GNLSE in the frequency space. An immediate consequence of these fundamental symmetries is that, if the initial condition is such that $A(0,T)=A^{*}(0,-T)$ (e.g., a transform limited pulse), then the forward and backward propagation directions are exactly equivalent. 

This result sets a limitation for the direct application of the reverse design method to soliton self-compression, where an ultrashort pulse propagating in the anomalous dispersion regime of a HCF can simultaneously broaden its spectrum and compensate its phase, due to the interplay between the opposite sign chirps introduced by the negative dispersion and SPM \cite{Travers2019,Brahms2020,Galan2022,Galan2023}. With the pulse compression taking place in a single stage, this scenario constitutes a simple starting point to test the reverse design procedure, as one can simply choose the desired compressed pulse and simulate its backward propagation towards the fiber entrance.

Ideally, we could think of finding a situation where an initial shaped pulse perfectly self-compresses to reach the shortest possible duration at the fiber output, i.e., the Fourier limit of the output spectrum. However, recalling the symmetries of the GNLSE, the forward and backward propagations are exactly equivalent for a transform limited pulse. Thus, as in nonlinear scenarios dominated by SPM Fourier limited pulses (FLPs) always broaden their spectrum during their forward propagation \cite{Agrawal}, just the same occurs in the backward direction, and this is true even if fiber losses and SRS are not negligible, because these effects do not alter the direction of the spectral broadening (see the Supplemental Material for a more detailed analysis on the symmetries of the GNLSE and the role of linear absorption and SRS). As a consequence, in a nonlinear regime dominated by SPM, an initial pulse can only self-compress to a FLP if it has a broader spectrum than the output, which is the opposite to what is sought in pulse compression experiments. In addition to not representing an actual compression scenario, the back-propagation of a FLP often results in an impractical complex input that would be difficult to shape in a real experiment, just the same as one suspects would happen when propagating it in the forward direction.

The above discussion raises an important constraint for the reverse propagation method. Namely, that in nonlinear scenarios typical of compression experiments, the desired output pulse cannot be chosen in a simple manner, but it must approach an actual nonlinear solution in both amplitude and phase. To overcome these limitations, we have introduced an intermediate pulse at the fiber output that can then be optimally post-compressed in a standard dispersive compression stage to produce the goal few-cycle pulse. In this work, we have tested the reverse design method with two different compressors: a standard set of pairs of diffraction gratings and a d-scan, which is commonly used for simultaneous few-cycle pulse compression and characterization \cite{Miranda2012,Bohle2014,Silva2014,Timmers2017}.

%--------------------------------------------------------
\section{Results and discussion}

\subsection{Reverse nonlinear propagation to find the\\ ideal pulse for post-compression}

\begin{figure}[htbp]
\includegraphics{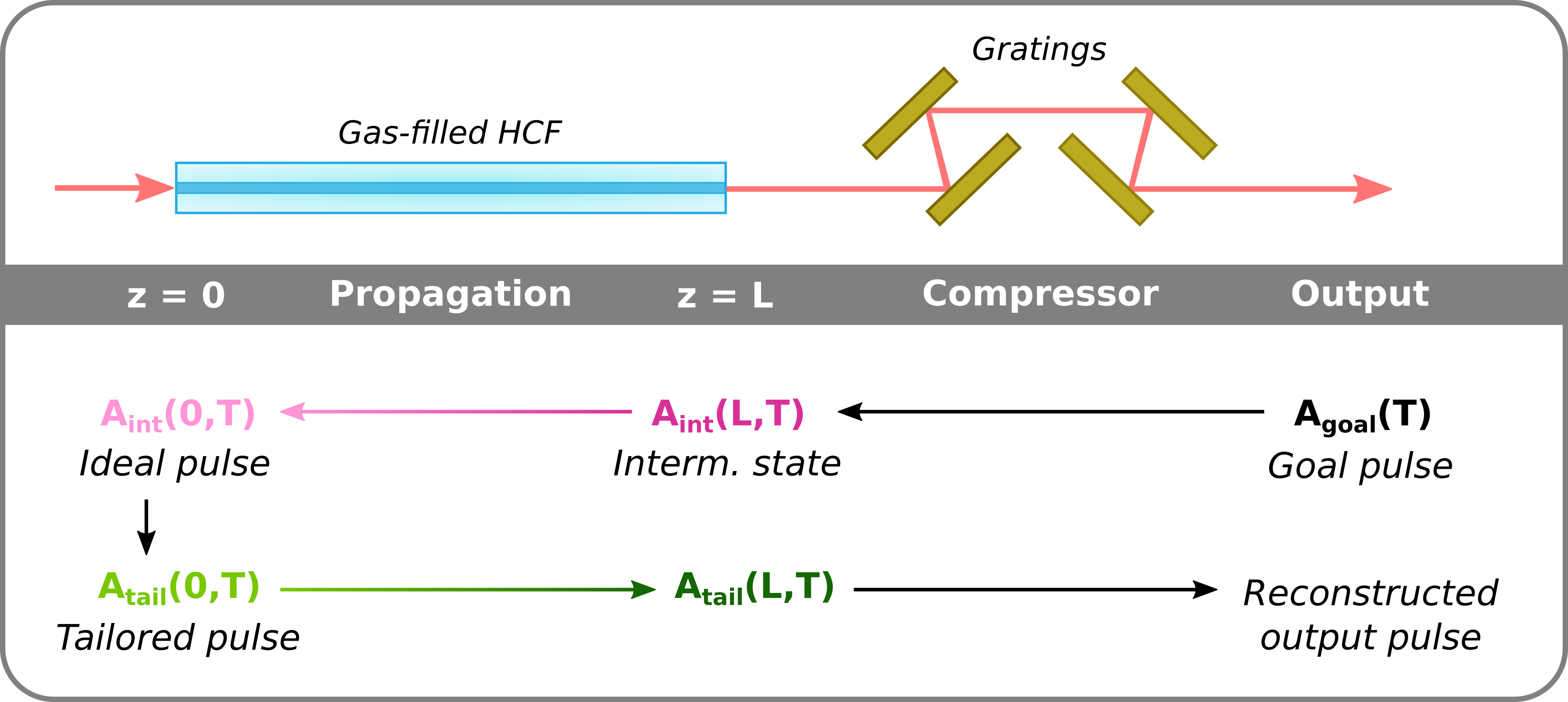}
\caption{\label{fig:esquema_redes} Flowchart describing the nonlinear reverse design method followed to generate ideal input pulses $A_{\mathrm{int}}(0,T)$ that could be optimally compressed to clean few-cycle goal pulses $A_{\mathrm{goal}}(T)$ after their propagation through a gas-filled HCF and a grating compressor. For clearness, the color code for the different pulses is preserved in all figures throughout the paper.}
\end{figure}

Our first proposal to test the reverse propagation method with a grating compressor is schematically depicted in the flowchart in Fig.~\ref{fig:esquema_redes}. We start by choosing a compressed goal pulse with a few-cycle Gaussian profile $A_{\mathrm{goal}}(T)=\sqrt{P_0}\exp[-T^2/(2T_0^2)]$ centered at $800$ nm, with an intensity full width at half-maximum (FWHM) duration of $8$ fs ($T_0=4.80$ fs) and a peak power $P_0=1.75$ GW. In this way, $A_{\mathrm{goal}}(T)$ models what could be the resulting ultrashort pulse from an ideal post-compression setup, without secondary temporal structures arising from uncompensated high-order dispersion terms remnant from the nonlinear interaction. This pulse is then reverse propagated through two pairs of gratings in a Treacy configuration which introduce a GDD of $-150.82$ fs$^2$ and TOD of $+197.21$ fs$^3$ \cite{Popmintchev2022}. As the compressor is assumed to be a linear stage, this simply chirps the goal pulse with the opposite phase to that introduced by the gratings, stretching it to a duration of $50.96$ fs and reducing its peak power to $0.27$ GW. The resulting chirped intermediate pulse $A_{\mathrm{int}}(L,T)$ is used for nonlinear reverse propagation through the 3 m long, 125 \textmu m core radius HCF filled with Ar at 1 bar. Note that, for this fiber, the zero-dispersion pressure is 0.47 bar, so at 1 bar the propagation takes place in the normal dispersion regime, as is usual in post-compression schemes. All pulse and fiber parameters were chosen --within standard and practicable experimental values-- to ensure the validity of the 1D-GNLSE model at the same time that the pulse accumulated enough nonlinear phase to broaden (narrow) its spectrum during forward (backward) propagation.

\begin{figure*}
\includegraphics[width=0.98\linewidth]{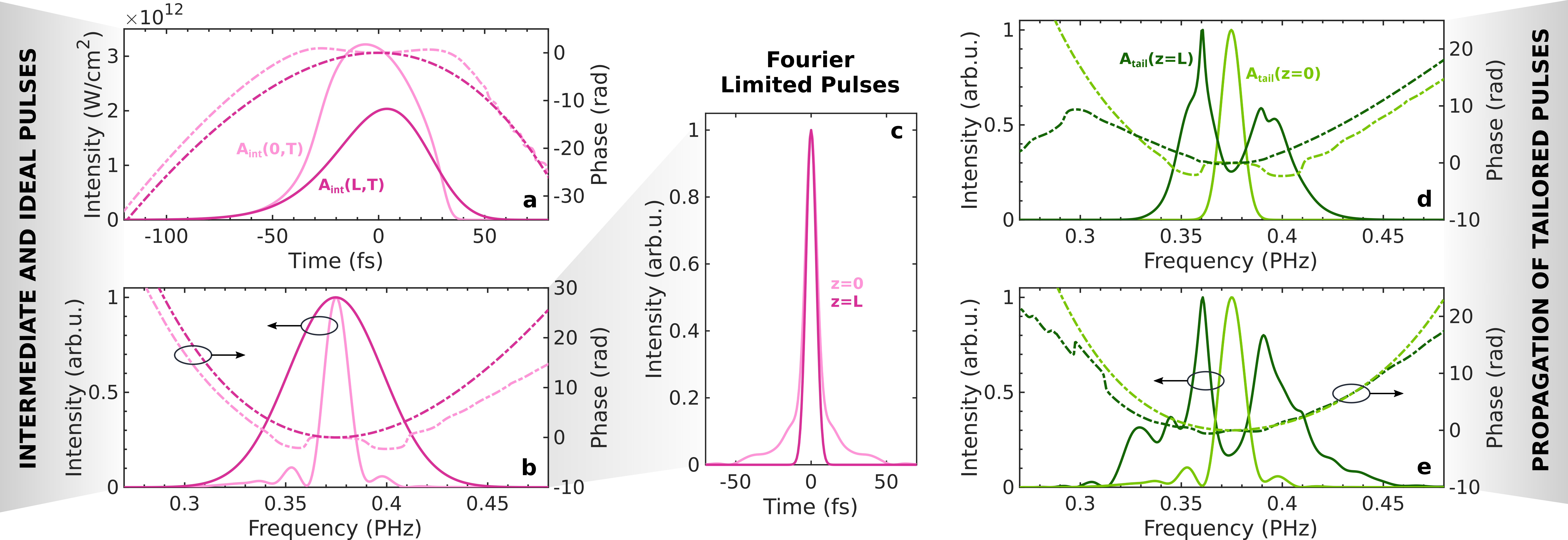}
\caption{\label{fig:direct} (a) Temporal and (b) spectral intensity profiles (solid line, left axis) and phase (dot-dashed line, right axis) at the entrance (light pink) and the end (fuchsia) of the HCF, corresponding to the ideal pulse that could be perfectly post-compressed to a goal few-cycle Gaussian pulse in a pair of gratings. (c) FLPs of the two spectra shown in (b). (d,e) Spectral intensity and phase of the tailored pulse at the fiber entrance (light green) and at the fiber output (dark green) after its forward nonlinear propagation. In (d) the tailored pulse was built by combining the ideal spectral phase and a Gaussian intensity fit, while in (e) it comprises the ideal spectral intensity and an approximated phase (see text for details).}
\end{figure*}

Figures~\ref{fig:direct}(a,b) show the temporal (a) and spectral (b) intensity and phase distributions of the chirped Gaussian intermediate pulse (fuchsia) an the resulting ideal pulse $A_{\mathrm{int}}(0,T)$ after reverse propagation (light pink). As we can see in Fig.~\ref{fig:direct}(b), the spectrum of the ideal pulse is slightly modulated, with an asymmetric spectral phase and an intensity profile with a main central peak and a pair of low-intensity side lobes. This secondary structure results from the mismatch between the chirp of $A_{\mathrm{int}}(L,T)$ and the nonlinear chirp induced mainly by SPM \cite{Boscolo2016}. Although the ideal spectrum at the HCF entrance is relatively complex, it seems to be narrower than the output one, which might be indicative of a proper post-compression scenario. It is interesting to note that this reverse spectral compression of the positively-chirped intermediate pulse is equivalent to that found in the conventional forward propagation of Gaussian-like pulses with negative chirp, where the initial linear frequency modulation is compensated for by the SPM-induced positive chirp \cite{Agrawal,Oberthaler1993,Washburn2000,Andresen2005,Finot2016,Daher2020}. The requirement that the intermediate Gaussian pulse must have positive GDD in order to narrow its spectrum in reverse propagation can be easily understood by recalling the GNLSE symmetries in the frequency domain as explained in Section~\ref{sec:simetrias}. However, it is straightforward to realize that this kind of spectral compression is deceiving. As we can see in Fig.~\ref{fig:direct}(c), the FLPs of the spectra at both HCF ends are almost identical, except for the pedestal structure that is present in the FLP of the input ideal pulse. Thus, this situation does not correspond to an effective compression process, but to a nonlinear reshaping of the pulse spectrum, increasing the energy carried around the central frequency, without introducing important changes in the corresponding FLP. Nevertheless, although this situation does not represent a compression scenario where the FLP duration is reduced, it can be seen as a reshaping stage to prepare an input pulse to undergo perfect compression.

When looking at the ideal spectral profile (light pink solid line in Fig.~\ref{fig:direct}(b)), it is natural to wonder how relevant are the low-intensity modulations that appear after the reverse propagation, and whether the main features of the ideal nonlinear propagation leading to the goal pulse could be recovered without them. To answer this question, we have fitted the modulated spectral intensity at the fiber entrance to a Gaussian distribution (light green solid line in Fig.~\ref{fig:direct}(d)) and combined it with the ideal spectral phase to build a new pulse, from now on referred to as tailored pulse $A_{\mathrm{tail}}(0,T)$, which closely resembles the ideal one. Finally, we have (forward) propagated the resulting pulse through the HCF under the same conditions as in the reverse propagation. To have an equivalent nonlinear interaction, we have forced the tailored pulse to have the same peak intensity than the ideal pulse. In Fig.~\ref{fig:direct}(d) we can see the spectrum at the fiber output obtained after the forward propagation of the tailored pulse, which turns out to be completely different from the initially proposed Gaussian profile. These results demonstrate that, even though, $A_{\mathrm{int}}(0,T)$ and $A_{\mathrm{tail}}(0,T)$ are very similar, their nonlinear propagation results in a completely different output situation, showing the great sensitivity of the nonlinearity to small changes in the initial spectral amplitude, and hindering the recovery of the perfectly-compressed goal pulse.

As a complementary test, we have analyzed what happens if we build the tailored pulse with the ideal spectral intensity and instead fit the spectral phase using a Taylor series around the central frequency up to a few dispersion orders. Figure~\ref{fig:direct}(d) shows the resulting output spectrum after forward propagating $A_{\mathrm{tail}}(0,T)$ comprising the ideal spectral amplitude and a spectral phase corresponding to a GDD of $+101.25$ fs$^2$, a TOD of $-205.44$~fs$^3$ and a fourth-order dispersion (FOD) of $+1.35\times10^3$ fs$^4$, an expansion that faithfully reproduces the ideal phase inside the main spectral lobe. Again, the output spectrum is quite different from the goal Gaussian distribution, proving that, in this situation, both the ideal phase and amplitude profiles are important to achieve a perfect pulse compression. We should note that similar results are obtained when using different goal pulse durations or different compressors. For the case of ultrashort goal pulses close to the single-cycle limit, the asymmetries and modulations of the ideal spectrum increase both in amplitude and phase, entering a very complex scenario which is out of the scope of this manuscript (see Supplemental Material for some examples).

\subsection{Reverse nonlinear propagation to improve\\ the post-compression process}

\begin{figure}[b]
\includegraphics{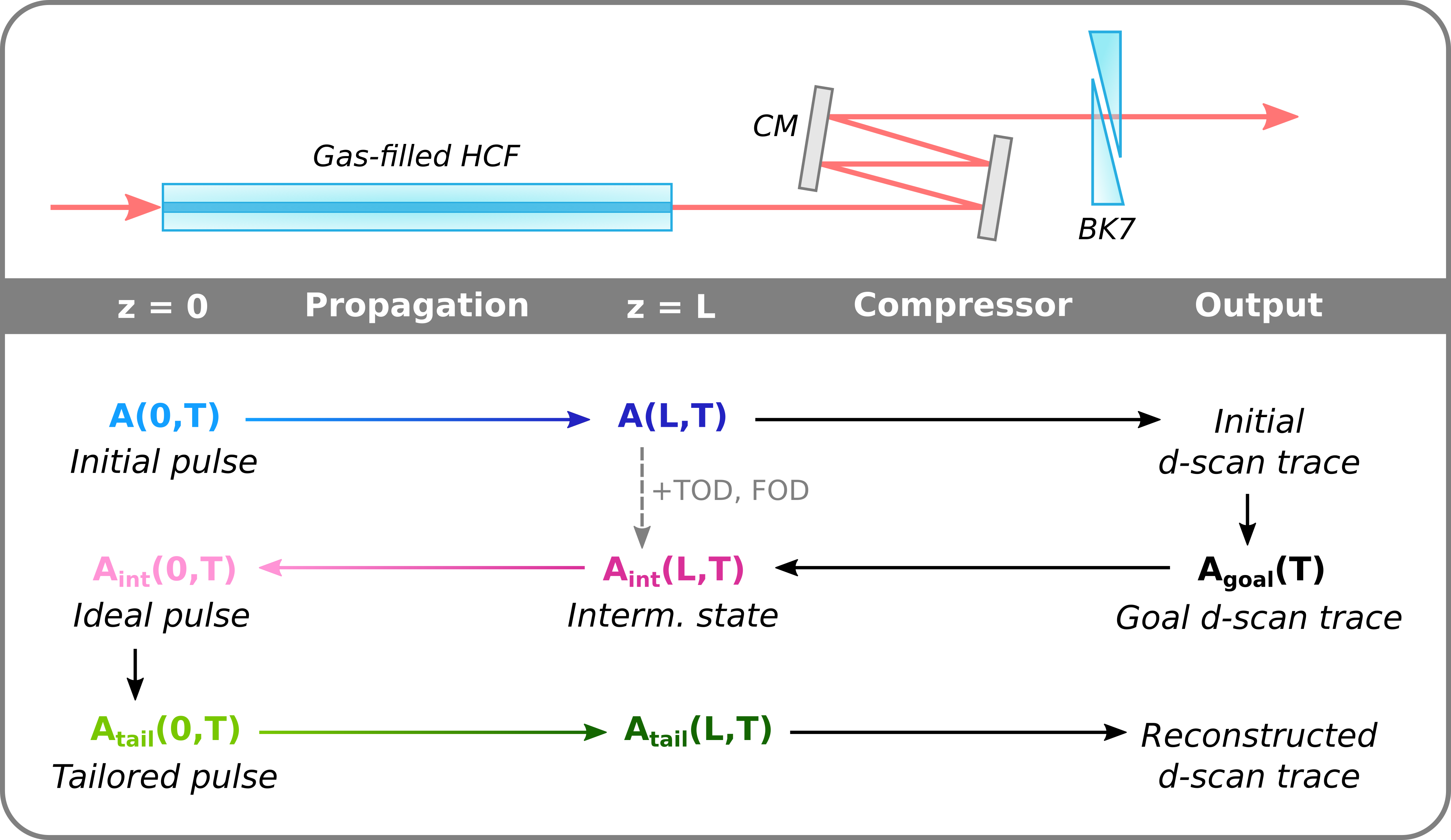}
\caption{\label{fig:esquema_dscan} Flowchart describing the nonlinear reverse design method followed to generate ideal shaped pulses $A_{\mathrm{int}}(0,T)$ that could be optimally compressed in a d-scan system after their nonlinear propagation through a gas-filled HCF.}
\end{figure}

\begin{figure*}
\includegraphics[width=0.95\linewidth]{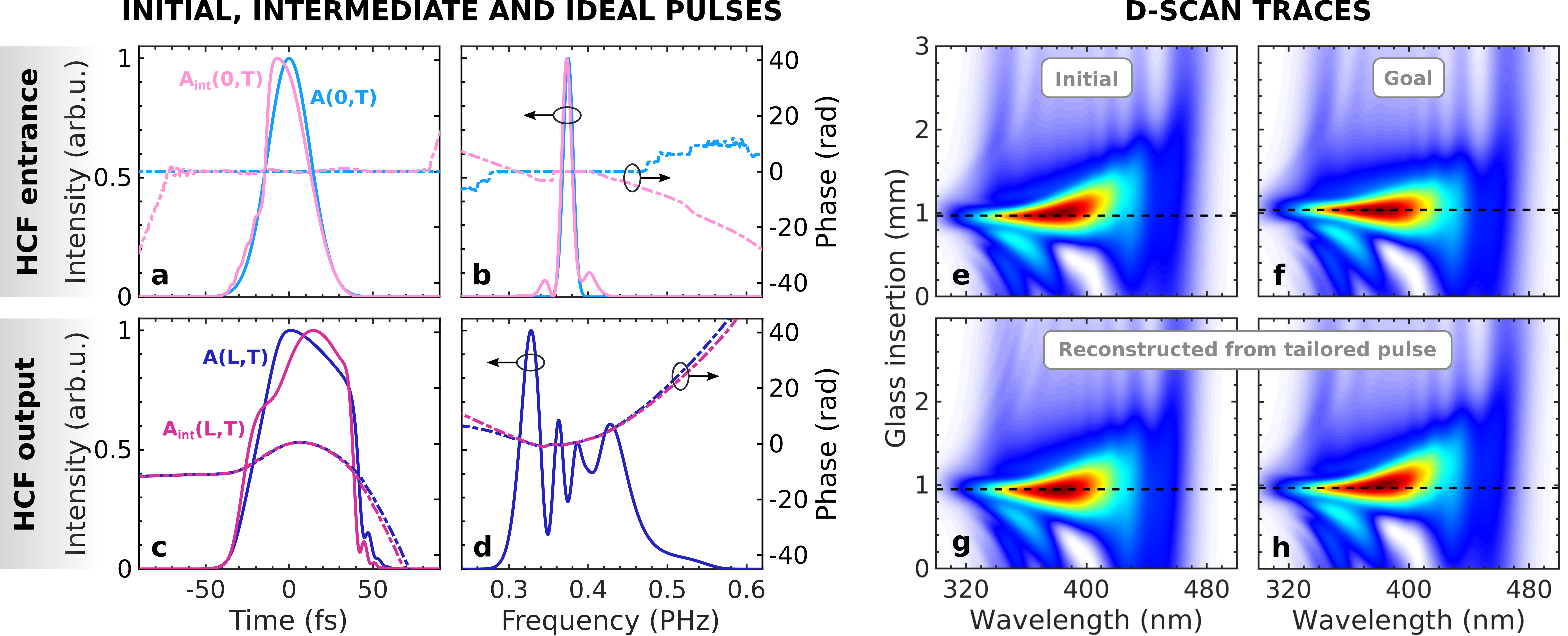}
\caption{\label{fig:casogauss} Reverse design of the ideal input pulse $A_{\mathrm{int}}(0,T)$ for optimal post-compression in a d-scan starting from an initial Gaussian pulse. (a) Initial pulse $A(0,T)$ and ideal pulse $A_{\mathrm{int}}(0,T)$ after back-propagation of the intermediate state, and (b) their corresponding spectra at the fiber entrance. (c) Output pulse $A(L,T)$ obtained after the forward propagation of the initial Gaussian pulse through the HCF, and intermediate pulse with optimal d-scan trace $A_{\mathrm{int}}(L,T)$ used as the initial condition for the reverse propagation. (d) Corresponding spectra at the fiber output. Solid lines are for intensity (left axis) and dot-dashed lines represent the temporal and spectral phases (right axis). (e) Initial d-scan trace of $A(L,T)$, (f) goal trace of $A_{\mathrm{int}}(L,T)$, and d-scan traces obtained after forward propagating the original Gaussian pulse with (g) the complete ideal spectral phase of $A_{\mathrm{int}}(0,T)$ and (h) an approximated input spectral phase up to FOD. The dashed horizontal lines over the d-scan traces represent the optimal glass insertion for maximum pulse compression.}
\end{figure*}

The previous section has shown that selecting a too-ideal goal pulse is not a feasible situation, both because it produces a complex input pulse that would be difficult to shape in a real experiment with enough precision so as to recover the ideal propagation, and because it does not lead in general to an effective post-compression scenario. Therefore, we now try to relax the output condition by choosing a goal pulse which is at the same time optimally compressed and close to an actual nonlinear solution, with a typical SPM-modulated spectrum and a complex nonlinear phase. In this section, we test the applicability of the nonlinear reverse propagation method to improve the results obtained beforehand from a standard HCF post-compression setup with a d-scan, which allows for a direct assessment of the post-compression quality and its optimization. The new proposal is schematically depicted in the flowchart of Fig.~\ref{fig:esquema_dscan}.

An initial simple pulse $A(0,T)$ is first propagated through a gas-filled HCF and its spectrum is broadened by SPM. The resulting output $A(L,T)$ is compressed in the d-scan module, and the compression performance is directly evaluated from the appearance of the d-scan trace. Intuitively, a flat and thin trace would be indicative of excellent compression, but actual d-scan traces often exhibit more complex tilts and curvatures which are a signature of residual higher-order dispersion terms from the nonlinear interaction \cite{Suda2012,Jarque2018}. Especially, TOD, which appears in the d-scan trace as a linear tilt, has proven to be one of the main limiting factors in many in-line post-compression experiments \cite{Silva2018,Bohle2014,Silva2014,Timmers2017}. The optimization step then consists in adding small amounts of TOD and FOD to $A(L,T)$ until its d-scan trace is optimally flattened (the goal d-scan trace). Finally, the intermediate phase-optimized pulse, $A_{\mathrm{int}}(L,T)$, is used as the initial condition to solve the nonlinear propagation equation backwards again towards the fiber entrance. This yields the ideal input pulse $A_{\mathrm{int}}(0,T)$ that should be launched in the HCF to achieve the best compressed ultrashort pulse after the d-scan. The way of building $A_{\mathrm{int}}(0,T)$ together with the reversibility of the GNLSE, ensures that the reverse propagation produces a practicable input. Our strategy is somehow similar to the hybridization proposed by Berti \textit{el al}. in the context of filamentation to ensure the collapse of the back-propagated target waveform into a filament \cite{Berti2015}.

In Fig.~\ref{fig:casogauss} we can see the results obtained for an initial transform limited Gaussian pulse $A(0,T)=\sqrt{P_0}\exp[-T^2/(2T_0^2)]$ centered at 800 nm, with a standard intensity FWHM duration of 30 fs ($T_0=18.02$ fs) and an input peak power $P_0=1.6$ GW (light blue line in Fig.~\ref{fig:casogauss}(a)). As before, these parameters were chosen to ensure both the validity of the 1D theoretical model and that the broadened spectrum by SPM could be compressed to a few-cycle pulse. After the nonlinear propagation of $A(0,T)$ through the same 3 m long, 125 \textmu m core radius HCF filled with Ar at 1 bar, the resulting output pulse (dark blue line in Fig.~\ref{fig:casogauss}(c)) has a d-scan trace with tilt and curvature which are a signature of the higher-order dispersion remnant from the nonlinear interaction (Fig.~\ref{fig:casogauss}(e)). In this case, the d-scan compressor was composed of ultra-broadband chirped mirrors introducing $-100$ fs$^2$ of pure GDD, and a pair of BK7 wedges. By adding the extra dispersion terms $\mathrm{TOD} = -30$ fs$^3$ and $\mathrm{FOD} = +60$ fs$^4$ to $A(L,T)$, its d-scan trace is optimized as seen in Fig.~\ref{fig:casogauss}(f). The resulting pulse $A_{\mathrm{int}}(L,T)$ corresponding to this goal trace (fuchsia line in Fig.~\ref{fig:casogauss}(c)) is finally back-propagated, yielding the ideal input pulse $A_\mathrm{int}(0,T)$ plotted in light pink in Fig.~\ref{fig:casogauss}(a). The latter is quite close to the original Gaussian pulse but, again, presents a modulated spectrum (light pink line in Fig.~\ref{fig:casogauss}(b)), with two low-intensity side lobes which are very similar to those found in the previous section.

\begin{figure}[b]
\includegraphics{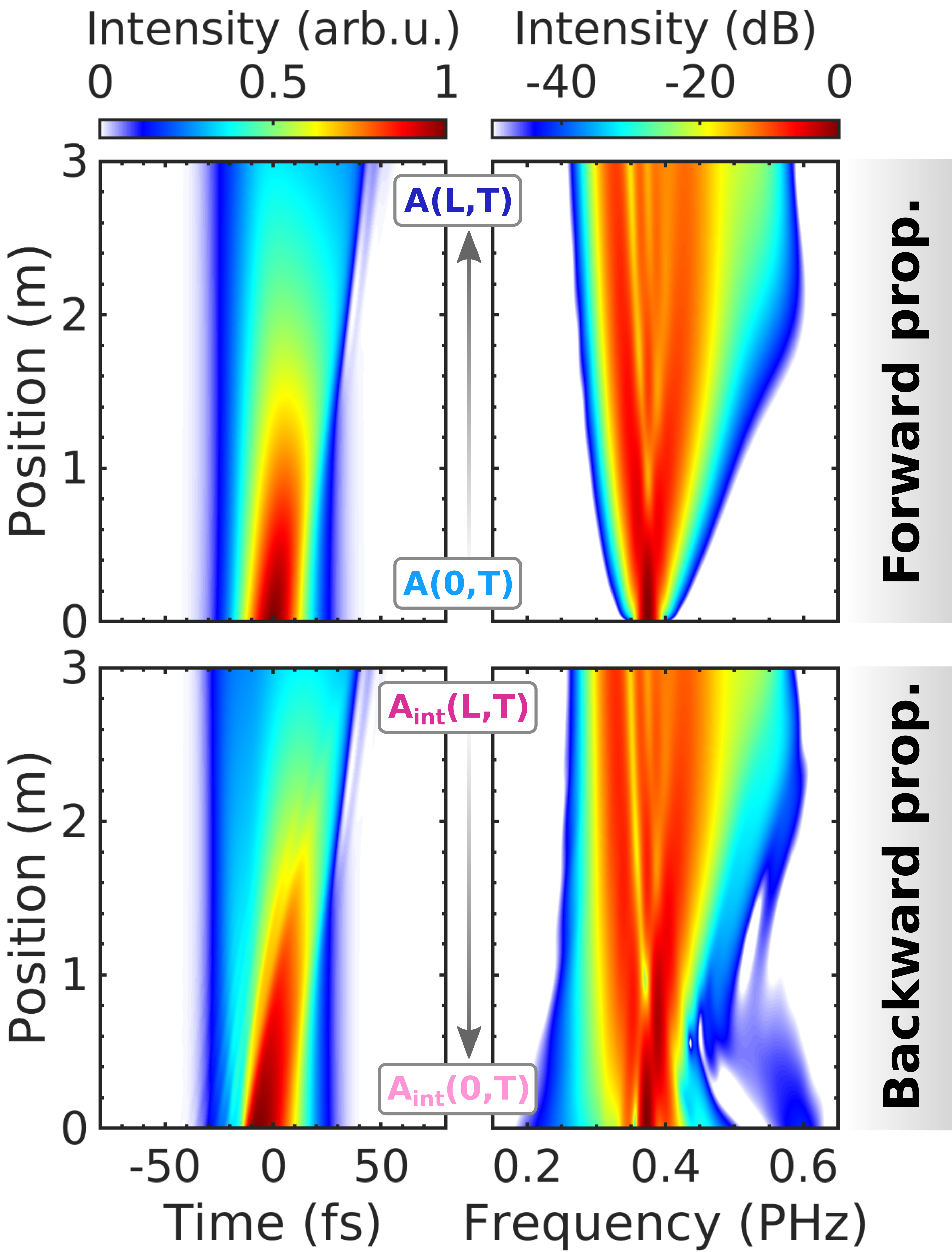}
\caption{\label{fig:evolutionGauss} Complete temporal (left) and spectral (right) evolution of the Gaussian initial pulse $A(0,T)$ (top row) and the corresponding $A_{\mathrm{int}}(L,T)$ (bottom row) during their forward and backward propagation through the HCF, respectively.}
\end{figure}

To illustrate the main features of the nonlinear propagation through the HCF in this scenario, Fig.~\ref{fig:evolutionGauss} shows the computed full temporal and spectral evolution of both $A(0,T)$ (top row) and $A_{\mathrm{int}}(L,T)$ (bottom row) during their forward and backward propagation, respectively. As we can see, in the forward direction, $A(0,T)$ broadens its spectrum by SPM and temporally disperses due to the combined positive chirp induced by SPM and the fiber normal dispersion. This is the typical evolution observed in most HCF-based post-compression  setups. The backward propagation of $A_{\mathrm{int}}(L,T)$ resembles the previous forward evolution, but close to the fiber entrance the pulse spectrum develops the characteristic side lobes.

As before, to find out if both the amplitude and spectral profiles of the ideal pulse are necessary to recover the goal d-scan trace in this second scenario, we have performed some approximations to $A_{\mathrm{int}}(0,T)$ and simulated its forward propagation. First, we have replaced the input shaped spectrum by the original Gaussian amplitude, neglecting the low-intensity side lobes. When the input tailored pulse generated by combining the Gaussian spectrum with the complete spectral phase of $A_{\mathrm{int}}(0,T)$ is forward propagated through the HCF and subsequently compressed in the d-scan, its trace (Fig.~\ref{fig:casogauss}(g)) slightly deviates from the optimized one but, surprisingly, it is still clearly better than the initial d-scan trace, showing that the optimization performed with the reverse design procedure can be partially recovered. Second, we have further simplified the ideal input pulse by approximating the predicted spectral phase of $A_{\mathrm{int}}(0,T)$ with a Taylor series expansion around the central frequency up to fourth order. This fit, which yields the dispersion terms $\mathrm{GDD}=-23.71$ fs$^2$, $\mathrm{TOD}=+74.90$ fs$^3$ and $\mathrm{FOD}=+8.555 \times 10^{3}$ fs$^4$, faithfully reproduces the theoretical phase inside the main peak of the Gaussian spectrum. The forward propagation of the original Gaussian spectrum with these dispersion orders results in the d-scan trace shown in Fig.~\ref{fig:casogauss}(h). In this case, the reconstructed trace is almost identical to the original one, and all the optimization performed with the reverse design method is lost by approximating the optimized input spectral phase. This fact reveals the extraordinary sensitivity of the nonlinearity to the phase of the initial condition, even in regions with low spectral amplitude.  

\begin{figure*}
\includegraphics[width=0.95\linewidth]{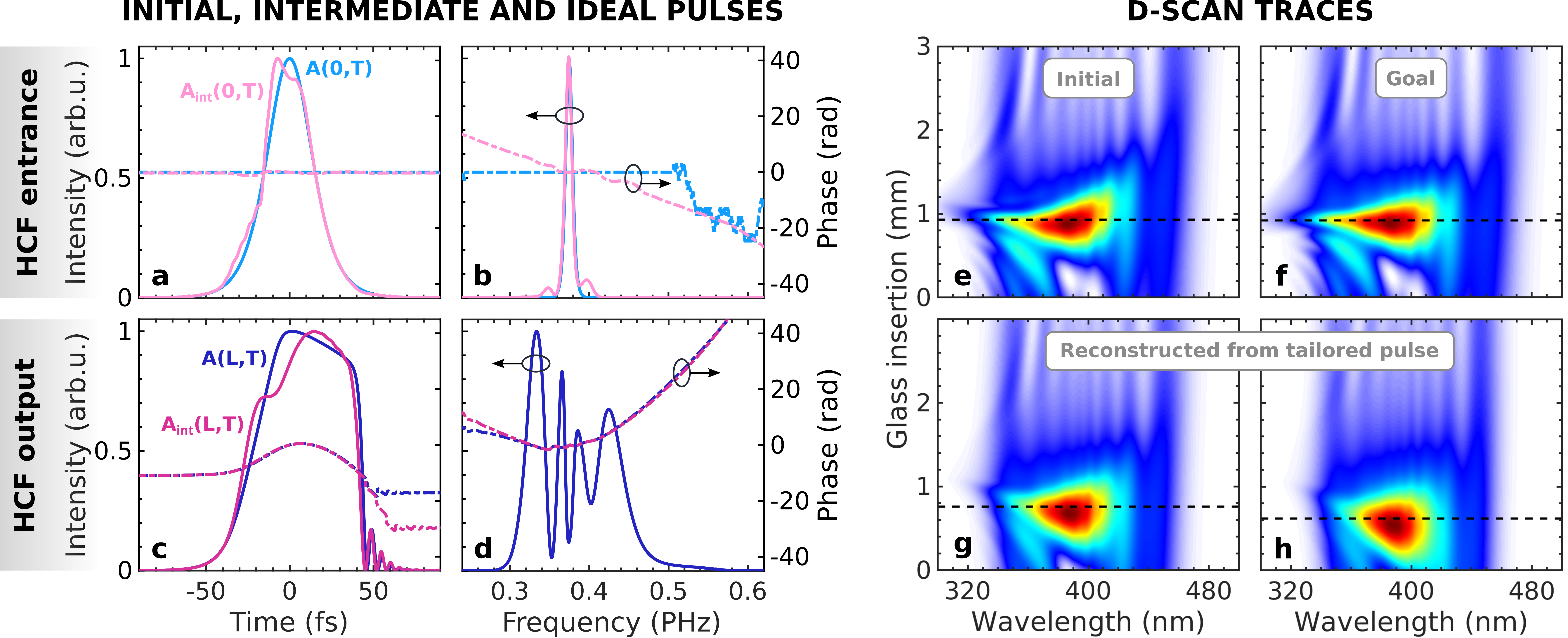}
\caption{\label{fig:casosech} Reverse design method of the ideal input pulse $A_{\mathrm{int}}(0,T)$ for optimal post-compression in a d-scan, applied to an initial hyperbolic secant pulse. (a) Initial pulse $A(0,T)$ and ideal pulse $A_{\mathrm{int}}(0,T)$ after nonlinear back-propagation of the intermediate state, and (b) their corresponding spectra at the fiber entrance. (c) Output pulse $A(L,T)$ obtained after the forward propagation of the initial sech pulse through the HCF, and intermediate pulse with optimal d-scan trace $A_{\mathrm{int}}(L,T)$ used as the initial condition for the reverse propagation. (d) Corresponding spectra at the fiber output. (e) Initial d-scan trace of $A(L,T)$, (f) goal trace of $A_{\mathrm{int}}(L,T)$, and d-scan traces obtained after forward propagating the original sech pulse with (g) the complete ideal spectral phase of $A_{\mathrm{int}}(0,T)$ and (h) an approximated input spectral phase up to FOD.}
\end{figure*}

Finally, to further illustrate the great sensitivity of the propagation to changes in the ideal pulse and to show how different conditions converge to a similar ideal spectrum in this second scenario, we have repeated the previous analysis starting from a slightly different initial pulse shape. In Fig.~\ref{fig:casosech} we can see the results obtained with the nonlinear reverse design method for an initial unchirped hyperbolic-secant pulse $A(0,T)=\sqrt{P_0}\sech(T/T_0)$ centered at 800 nm, with an intensity FWHM duration of 31.76 fs ($T_0=18.02$ fs) and an input peak power $P_0=1.6$ GW (light blue line in Fig.~\ref{fig:casosech}(a)). All pulse parameters have been chosen so that the dispersion and nonlinear lengths match those of the previous Gaussian case for the same 3 m long, 125 \textmu m HCF filled with 1 bar of Ar ($L_{D} = 5.58$ m and $L_{NL} = 0.17$ m). In this way, both situations can be fairly compared. For the usual definitions of the characteristic lengths $L_D$ and $L_{NL}$ the reader is referred to \cite{Chen2002,Agrawal,Travers2011}.

As we can see in Figs.~\ref{fig:casosech}(a-d), all the pulses at both the fiber entrance and exit are very close to the previous results for a Gaussian pump. In this case, the initial d-scan trace was optimized by adding the almost identical dispersion terms $\mathrm{TOD}=-30$ fs$^3$ and $\mathrm{FOD}=+100$ fs$^4$ to $A(L,T)$. Furthermore, the spectrum of the ideal input pulse $A_{\mathrm{int}}(0,T)$ now exhibits smaller side lobes and greatly approaches the original sech pulse spectral amplitude. Thus, one would expect that, in this case, the approximations performed to the ideal input shape should allow for a better recovery of the goal d-scan trace. However, just the opposite happens. As we can see in Fig.~\ref{fig:casosech}(g), now the mere substitution of the modulated spectrum of $A_{\mathrm{int}}(0,T)$ by the initial sech amplitude results in a clearly deteriorated d-scan trace after the nonlinear forward propagation of the tailored pulse though the HCF. The situation further worsens when the input spectral phase is also replaced by a Taylor series expansion ($\mathrm{GDD}=+57.45$ fs$^2$, $\mathrm{TOD}=+1.204 \times 10^3$ fs$^3$ and $\mathrm{FOD}=+1.735 \times 10^5$ fs$^4$) as shown in Fig.~\ref{fig:casosech}(h). All these results reveal the intricate sensitivity of the nonlinear propagation equation to the initial condition, suggesting that, contrary to what is usually assumed in ultrashort pulse post-compression experiments with phase pre-compensation \cite{Silva2018,Silva2014,Suda2012}, in general both the input pulse phase and amplitude must be shaped --to a greater or lesser extent depending on the specific situation-- in order to achieve optimal compression. This behavior has been recently pointed out in \cite{Hergott2023}, where both experiments and numerical simulations demonstrated the extreme sensitivity of the nonlinear spectral broadening and post-compressed pulse duration to the initial bandwidth and phase of the pump pulse.

%--------------------------------------------------------
\section{Conclusions}

We have demonstrated that the reverse nonlinear propagation method can be applied to find the ideal pulse structure that yields a desired ultrashort and clean goal pulse, after its propagation through a HCF-based post-compression setup. We have shown that, in typical nonlinear scenarios dominated by SPM, the reverse design method requires a chirped intermediate state in order to overcome the limitations imposed by the fundamental symmetries of the propagation equation and generate a feasible input that could be shaped in a standard experiment. In this work, we have tested the applicability of the reverse propagation method to predict the exact input pulse shape that could lead either to an ideal few-cycle post-compressed pulse, or to an improved compression scenario in a standard HCF setup with a d-scan-based compression stage. Two common conclusions are obtained from our analysis: first, that the ideal input pulses typically present a modulated spectrum with characteristic low-intensity side lobes. And second, our results reveal the intricate sensitivity of the nonlinear propagation equation to the initial condition showing that, in general, both input phase and amplitude shaping may be necessary to achieve the desired optimal compression, as both quantities are inevitably linked by the nonlinearity.

\begin{acknowledgments}
We thank Professors L. Plaja and I. J. Sola for fruitful comments and discussions. This project has received funding from Ministerio de Ciencia e Innovación (MCIN/AEI/10.13 039/501100011033, I+D+i grant PID2019-106910GB-I00 and PID2022-142340NB-I00). M.F.G. acknowledges support from Ministerio de Universidades under grant FPU21/02916.
\end{acknowledgments}

% The \nocite command causes all entries in a bibliography to be printed out
% whether or not they are actually referenced in the text. This is appropriate
% for the sample file to show the different styles of references, but authors
% most likely will not want to use it.
% \nocite{*}

%\bibliography{references}

%apsrev4-2.bst 2019-01-14 (MD) hand-edited version of apsrev4-1.bst
%Control: key (0)
%Control: author (8) initials jnrlst
%Control: editor formatted (1) identically to author
%Control: production of article title (0) allowed
%Control: page (0) single
%Control: year (1) truncated
%Control: production of eprint (0) enabled
%

\newpage

\begin{figure*}
\includegraphics[width=\linewidth]{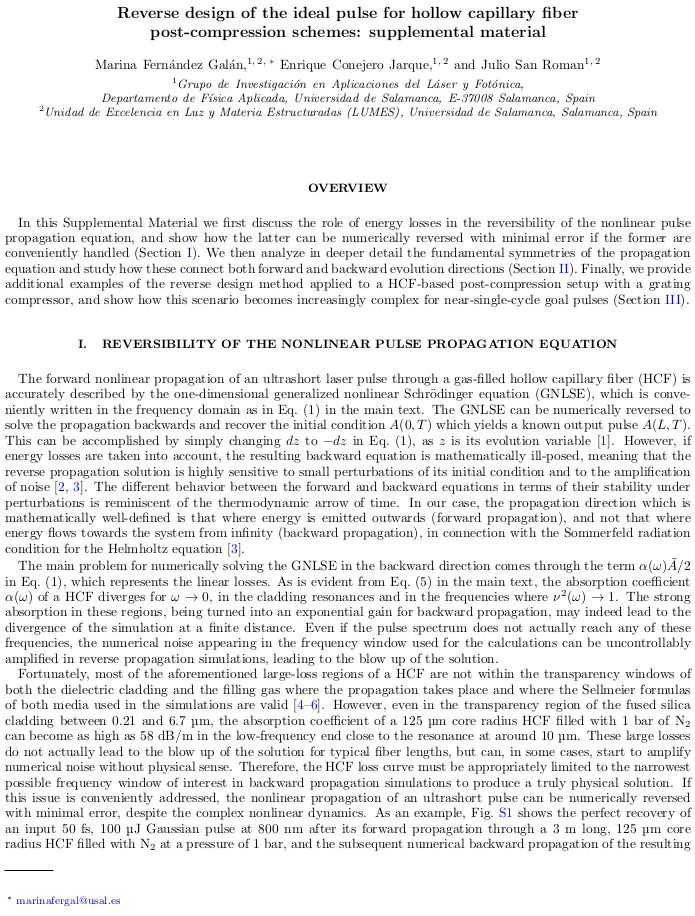}
\end{figure*}

\begin{figure*}
\includegraphics[width=\linewidth]{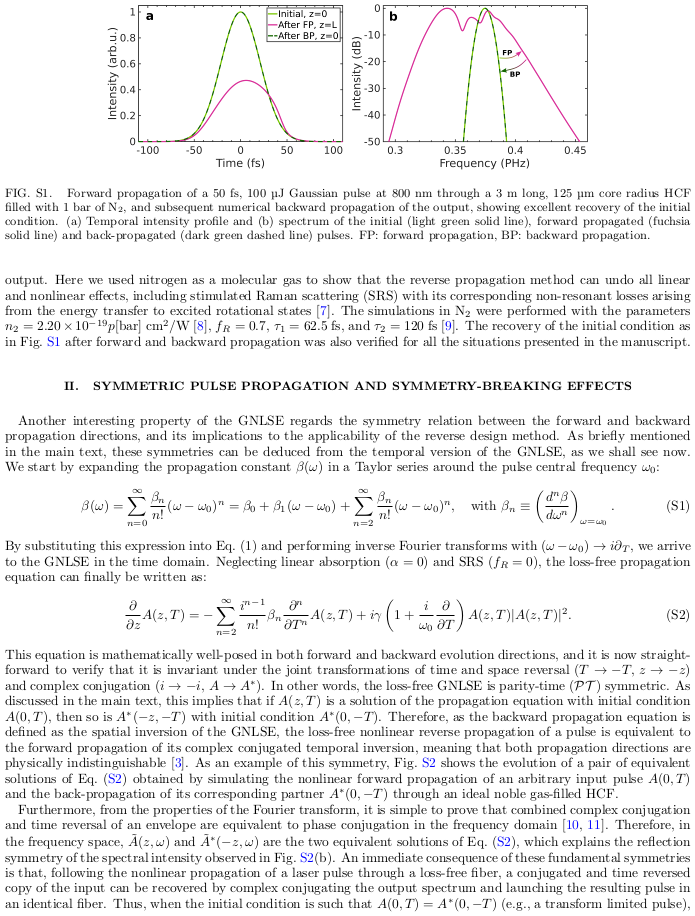}
\end{figure*}

\begin{figure*}
\includegraphics[width=\linewidth]{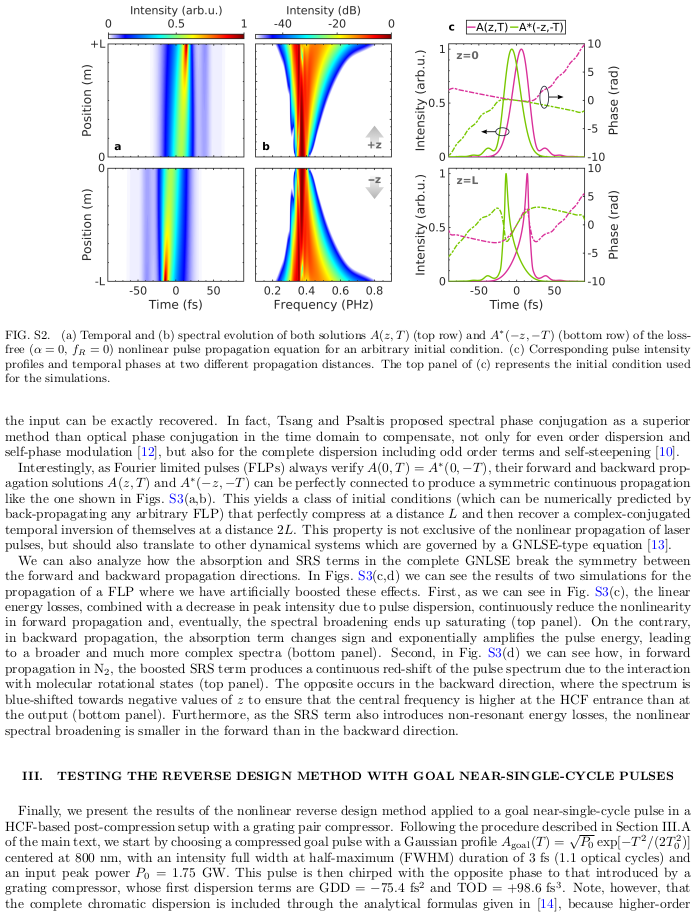}
\end{figure*}

\begin{figure*}
\includegraphics[width=\linewidth]{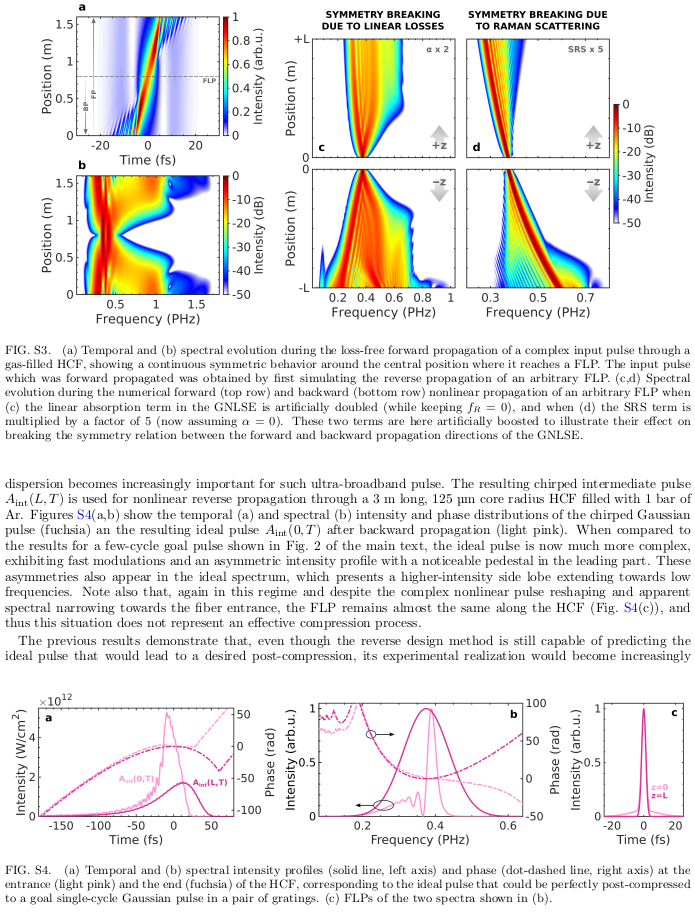}
\end{figure*}

\begin{figure*}
\includegraphics[width=\linewidth]{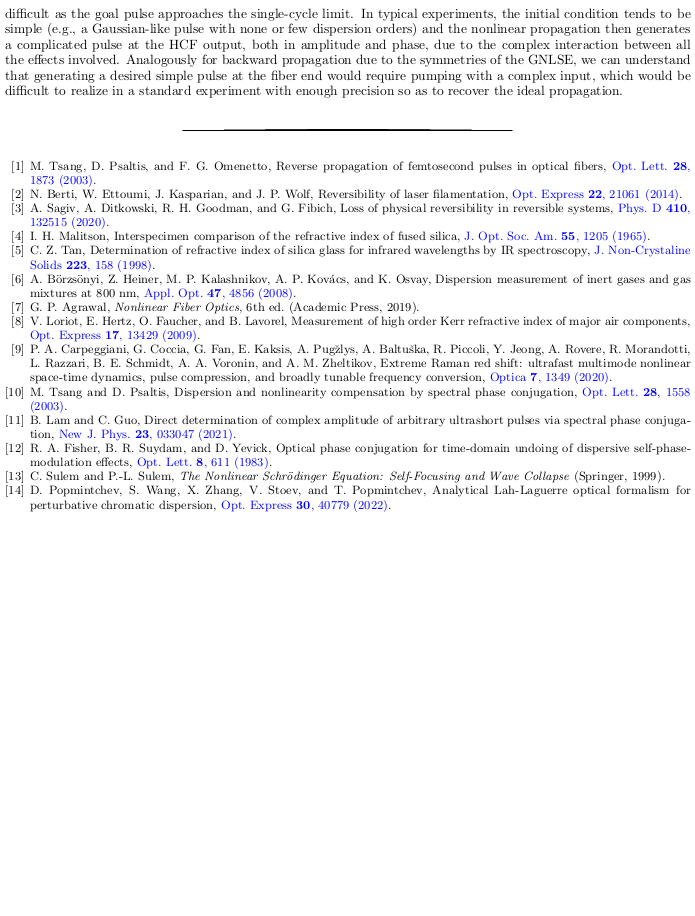}
\end{figure*}

\end{document}